\documentclass[prd, preprint, nofootinbib]{revtex4}
\usepackage{epsfig,graphicx,amsmath,amssymb}
\usepackage{graphicx}
\usepackage{amsfonts}
\usepackage{latexsym}
\usepackage{amssymb}
\usepackage{slashed}

\usepackage{epstopdf}
\usepackage[bookmarks=true,bookmarksnumbered=true,bookmarkstype=toc, colorlinks=true, 
citecolor=blue, linkcolor=blue ,urlcolor=blue]{hyperref} 
\usepackage{units}

\usepackage{color}

\preprint{UME-PP-021}
\preprint{EPHOU-22-013} 
\preprint{KYUSHU-HET-244}

\begin{document}
                 
\title{Revisiting sterile neutrino dark matter in gauged $U(1)_{B-L}$ model}

\author{Shintaro Eijima}
\email{eijima@icrr.u-tokyo.ac.jp}
\affiliation{ICRR, The University of Tokyo, Kashiwa, Chiba 277-8582, Japan}

\author{Osamu Seto}
 \email{seto@particle.sci.hokudai.ac.jp}
\affiliation{Department of Physics, Hokkaido University, Sapporo 060-0810, Japan}

\author{Takashi Shimomura}
\email{shimomura@cc.miyazaki-u.ac.jp}
\affiliation{Faculty of Education, Miyazaki University, Miyazaki 889-2192, Japan}
\affiliation{Department of Physics, Kyushu University, Fukuoka 819-0395, Japan}

%
\begin{abstract}
We reexamine sterile neutrino dark matter in gauged $U(1)_{B-L}$ model.  
Improvements have been made by tracing and careful evaluation of the evolution of the number densities of sterile neutrinos $N$ and extra neutral gauge bosons $Z'$.
As a result, the cosmologically-interesting gauge coupling of $U(1)_{B-L}$ for freeze-in sterile neutrinos turns out to be smaller than the values reported in the literature.
This avoids the overproduction of $Z'$ so that it is consistent with the big bang nucleosynthesis and the cosmic microwave background constraints on the effective number of neutrino species.
Similarly, the free-streaming length constraints exclude a large parameter space derived in previous studies.
In addition to known freeze-in pair production of $N$ from the standard model fermion pairs,
we find the case that $N$ is dominantly produced from a pair of $Z'$ at the temperature characterized by the $B-L$ breaking scalar mass. 
Thus, the naive truncation of the $U(1)_{B-L}$ scalar contribution made in the literature is not valid.
\end{abstract}

\vspace*{1cm}
\maketitle


\section{Introduction}

Not only the existence of nonbaryonic dark matter (DM) but also nonvanishing neutrino masses are major open questions in particle physics and cosmology. 
Those are also hints and evidence for new physics beyond the Standard Model (SM). 
One of the simple extensions of the SM to explain those two problems is to introduce right-handed (RH) neutrinos, which are singlets under the SM gauge group.
With Majorana masses of RH neutrinos, the tiny neutrino masses can be naturally generated through the seesaw mechanism~\cite{Minkowski:1977sc,Yanagida:1979as,GellMann:1980vs,Mohapatra:1979ia}.
The resultant heavier-mass eigenstates compared with the active neutrinos responsible for neutrino oscillations are called sterile neutrinos. They slightly mix with the left-handed (LH) components via the active-sterile mixings.
The lightest of them is a good candidate for (warm) DM~\cite{Dolgov:2000ew} if the lifetime is long enough~\cite{Drewes:2016upu,Boyarsky:2018tvu}.
The $\nu$MSM~\cite{Asaka:2005an,Asaka:2005pn} is known as an extension with three RH neutrinos including such a sterile neutrino dark matter.

Three generations of RH neutrinos can be theoretically verified, once the SM is extended by introducing an extra $U(1)$ gauge symmetry under which RH neutrinos are charged. 
Since RH neutrinos are chiral, their charges and generations are determined by gauge anomaly cancellation.
A well-studied example is gauged $U(1)_{B-L}$ symmetry~\cite{Davidson:1978pm,Mohapatra:1980qe,Marshak:1979fm}.

If RH neutrinos interact through an extra $U(1)$ gauge interaction, the sterile neutrino DM can be generated effectively. 
The Dodelson-Widrow (DW) mechanism~\cite{Dodelson:1993je}, where sterile neutrino DM is produced through the active-sterile mixings, conflicts with the observations in x-ray~\cite{Boyarsky:2005us,Boyarsky:2006fg,Boyarsky:2006ag,Boyarsky:2007ay,Yuksel:2007xh} and for Lyman-$\alpha$ forest (See, {\it e.g.}, \cite{Boyarsky:2018tvu} and references therein). 
Thus, the production from scatterings through new mediators~\cite{Khalil:2008kp,Kaneta:2016vkq,Biswas:2016bfo,Seto:2020udg,DeRomeri:2020wng,Lucente:2021har,Belanger:2021slj} is an alternative promising scenario.
This class of nonthermal production is sometime called ``freeze-in''~\cite{Hall:2009bx}. 
For a review, see, e.g., Refs.~\cite{Baer:2014eja,Shakya:2015xnx}. 
Nonthermal productions of an extra $U(1)$ gauge interacting sterile neutrino DM have been investigated for a large gauge coupling~\cite{Khalil:2008kp}, and for a small gauge coupling in Refs.~\cite{Kaneta:2016vkq,Biswas:2016bfo,FileviezPerez:2019cyn,Heeba:2019jho,Okada:2020cue,Okada:2021nwo,Iwamoto:2021fup}.

The purpose of this paper is to reexamine the freeze-in production of $U(1)_{B-L}$ gauge-interacting sterile neutrino DM.
After we examine the production processes of the sterile neutrino $N$ and the extra neutral gauge boson $Z'$ in detail and trace the evolution of the number densities of $N$ and $Z'$, 
 we evaluate cosmological constraints on the model from the viewpoint of light degrees of freedom and the structure formation.
We find that the inverse decay production of $Z'$ from SM particles in equilibrium is dominant in cases with a very small $U(1)_{B-L}$ gauge coupling.
Due to the contribution, the resultant abundances of the sterile neutrino DM and light $Z'$ as dark radiation (DR) are larger than previously estimated.
Therefore, the impacts of this increase on the big bang nucleosynthesis (BBN) and the cosmic microwave background (CMB) are estimated, in which constraints on the gauge coupling are obtained.
We also find that the contribution from the on shell $Z'$ decay is dominant in the scattering production process through $s$-channel $Z'$ exchange.
Since the produced $N$ from the $Z'$ decay is relativistic, the free-streaming length constraints limit the viable parameter space. 
Moreover, we find cases where the sterile neutrino DM is dominantly produced by the pair production from a pair of $Z'$.
The cross section of this mode depends on the mass of $U(1)_{B-L}$ breaking scalar. Thus, the naive truncation of the scalar contribution made in literature cannot be validated.

This paper is organized as follows. We describe Langrangian of the model, and read vertices of interactions in Sec.~\ref{sec:model}. 
After we summarize Boltzmann equations which need to be solved in Sec.~\ref{sec:bolt}, we consider three production scenarios of the 
sterile neutrino DM and estimate the DM abundance by taking into account cosmological constraints in Sec.~\ref{sec:dm}. 
We shortly discuss the implication to dark matter detection in Sec.~\ref{sec:detection}.
We discuss the conclusion in Sec.~\ref{sec:concl}.

\section{Model}
\label{sec:model}

We consider the extension of the SM by gauging $U(1)_{B-L}$ symmetry, where $B$ and $L$ are the baryon and 
lepton number, respectively. 
Under the $SU(3)_C\times SU(2)_L \times U(1)_Y \times U(1)_{B-L}$ gauge groups, three generations of RH neutrinos ($\nu_R^i$ with $i$ 
running over $1, 2, 3$) have to be introduced for the anomaly cancellation. 
The scalar sector of the SM is also extended by introducing one complex scalar $\Phi_{B-L}$, which is charged under the $U(1)_{B-L}$, to break the extra gauge symmetry spontaneously. 
This is the minimal extension regarding the gauged $U(1)_{B-L}$ symmetry. 
In Table~\ref{tab:matter-contents}, $Q$ $(u_R,~d_R)$ and $L$ $(e_R,~\nu_R)$ denote the LH (RH) quarks and leptons, 
respectively.

\begin{table}[htb]
	\centering
	\begin{tabular}{|c|ccc|c|} \hline	
		 & $SU(3)_C$ & $SU(2)_L$ & $U(1)_Y$  & $U(1)_{B-L} $ \\ \hline
		$Q^i$ & $\mathbf{3}$ & $\mathbf{2}$ & $\frac{1}{6}$ & $\frac{1}{3}$ \\
		$u_R^i$ & $\mathbf{3}$ & $\mathbf{1}$ & $\frac{2}{3}$ & $\frac{1}{3}$ \\
		$d_R^i$ & $\mathbf{3}$ & $\mathbf{1}$ & $-\frac{1}{3}$ & $\frac{1}{3}$ \\ \hline
		$L^i$ & $\mathbf{1}$ & $\mathbf{2}$ & $-\frac{1}{2}$ & $-1$ \\
		$e_R^i$ & $\mathbf{1}$ & $\mathbf{1}$ & $-1$ & $-1$ \\
		$\nu_R^i$ & $\mathbf{1}$ & $\mathbf{1}$ & $0$ & $-1$ \\ \hline
            $\Phi_H$ & $\mathbf{1}$ & $\mathbf{2}$ & $\frac{1}{2}$ & $0$ \\
            $\Phi_{B-L}$ & $\mathbf{1}$ & $\mathbf{1}$ & $0$ & $2$ \\ \hline
	\end{tabular}
\caption{
In addition to the SM particle content, three RH neutrinos  
 $\nu_R^i$ ($i=1, 2, 3$) and one $U(1)_{B-L}$ Higgs field $\Phi_{B-L}$ are introduced.   
}
\label{tab:matter-contents}
\end{table}

\subsection{Lagrangian}
The Lagrangian is given by
\begin{align}
 \label{LagF}
 \mathcal{L} &= \mathcal{L}_\text{SM} + \mathcal{L}_{\nu_R} + V(\Phi_H,\Phi_{B-L}), \\
 \label{Lag1} 
 \mathcal{L}_{\nu_R} &= \overline{\nu_R^i} (i D_\mu \gamma^\mu) \nu_R^i - y_{\nu}{}_{ij}\overline{L^{i}} \tilde{\Phi}_H \nu_R^j - \frac{1}{2}y_{\nu_R^i} \Phi_{B-L} \overline{\nu_R^{i~C}} \nu_R^i  + \mathrm{ H.c.} ,
\end{align}
where $\mathcal{L}_\text{SM}$, $\mathcal{L}_{\nu_R}$, and $V(\Phi_H,\Phi_{B-L})$ are Lagrangian of the SM, RH neutrinos, and the scalar potential of this model, respectively.
In Eq.~(\ref{Lag1}), the superscript $C$ denotes the charge conjugation of $\nu^i_R$, and $\tilde{\Phi}_H \equiv \epsilon \Phi_H^{\dagger}$ is the conjugation of the Higgs field $\Phi$ with $\epsilon$ being the antisymmetric tensor.
Yukawa couplings with LH lepton doublets and among RH neutrinos are denoted by $y_{\nu}$ and $y_{\nu_R^i}$ respectively, in which $i$ and $j$ are indices of flavor or generation.
We work on the diagonal basis of $y_{\nu_R^i}$ without loss of generality.

Due to the $U(1)_{B-L}$ symmetry, the covariant derivative is modified as 
\begin{align}
D_\mu = \partial_\mu - i g_2 W_\mu - i g_1 Y B_\mu - i g_{B-L} q_{B-L} X_\mu ,
\end{align}
where $W,~B$, and $X$ represent the gauge fields,  and 
$g_2,~g_1$, and $g_{B-L}$ are the gauge coupling constants of $SU(2)_L,~U(1)_Y$, and $U(1)_{B-L}$, respectively.  
The $U(1)_Y$ and $U(1)_{B-L}$ charges listed in Table~\ref{tab:matter-contents} are denoted as $Y$ and $q_{B-L}$.
We omit any symbol about the $SU(3)_C$ color interaction. 

There may exist a gauge kinetic mixing term
\begin{align} 
\mathcal{L}_{\varepsilon} = \frac{\varepsilon}{2} B_{\mu\nu} X^{\mu\nu},
\end{align}
where $B_{\mu\nu}$ and $X_{\mu\nu}$ are the gauge field strength of $U(1)_Y$ and $U(1)_{B-L}$ gauge field respectively, and $\varepsilon$ is a mixing parameter. 
The importance of this term has been studied intensively in the context of the so-called dark photon. Since we are interested in the $U(1)_{B-L}$ gauge interaction on RH neutrinos, we concentrate on the cases where the effect of the gauge kinetic mixing is negligible and set $\varepsilon$ vanishing in this paper.\footnote{If the gauge kinetic mixing effects are more dominant than the direct gauge interaction, the model would reduce to a dark photon model. For a recent review on the dark photon, see, e.g., Refs.~\cite{Caputo:2021eaa}.}

\subsection{Scalar potential}
\subsubsection{Mass eigenstates and Higgs mixing}
We derive the masses of introduced particles with the following scalar potential,
\begin{align}
V(\Phi_H, \Phi_{B-L}) = & \frac{1}{2} \lambda_1 \left( |\Phi_H|^2-\frac{v^2}{2}\right)^2+\frac{1}{2} \lambda_2 \left( |\Phi_{B-L}|^2-\frac{v_{B-L}^2}{2}\right)^2 \nonumber \\
 & +\lambda_3 \left(|\Phi_H|^2-\frac{v^2}{2}\right) \left( |\Phi_{B-L}|^2-\frac{v_{B-L}^2}{2}\right) ,
\label{eq:V}
\end{align}
where all parameters, $\lambda_1, \lambda_2, \lambda_3, v_{B-L}$, and $v \simeq 246$ GeV, are real and positive.
At the electroweak and $B-L$ broken vacuum, $\Phi_H$ and $\Phi_{B-L}$ fields can be expanded around those vacuum expectation values (VEVs) $v$ and $v_{B-L}$, respectively.

In this vacuum, the $U(1)_{B-L}$ gauge boson $X$ absorbs the corresponding Nambu-Goldstone  mode and becomes the massive extra neutral gauge boson $Z'$ with the mass 
\begin{align}
m_{Z'}^2 = 4 g_{B-L}^2 v_{B-L}^2, 
\label{eq:Z'mass}
\end{align}
and three RH neutrinos also obtain their Majorana masses
\begin{align}
m_{\nu_R^i} = \frac{y_{\nu_R^i} }{\sqrt{2}} v_{B-L}.
\end{align}
Then, the nonvanishing neutrino masses can be generated through the type-I seesaw mechanism as
\begin{align}
m_{\nu}{}_{ij} &=-m_D{}_{i k}\frac{1}{m_{\nu_R^k}}m_D^T{}_{k j}, 
\end{align}
with 
\begin{align}
 m_D{}_{i k} & = \frac{y_{\nu i k} v}{\sqrt{2}}.
\end{align}

The mass eigenstates of the scalars, $h$ and $\phi$, are obtained from those physical fluctuations $\phi_H$ and $\phi_{B-L}$ in $\Phi_H$ and $\Phi_{B-L}$, respectively, as
\begin{align}
&
\left( 
\begin{array}{c} 
\phi_H \\
\phi_{B-L} \\
\end{array}
\right)
 =
\left(
\begin{array}{cc}
 \cos\alpha & \sin\alpha  \\
 -\sin\alpha & \cos\alpha \\
\end{array}
\right)
\left( 
\begin{array}{c} 
h \\
\phi \\
\end{array}
\right) , 
\end{align}
with the mixing angle $\alpha$. 
Their masses are given by 
\begin{subequations}
\begin{align}
& m_h^2 = \frac{1}{2}\left( \lambda_1 v^2+\lambda_2 v_{B-L}^2 + \frac{\lambda_1 v^2-\lambda_2 v_{B-L}^2}{\cos(2 \alpha )} \right) , \\
& m_{\phi}^2 = \frac{1}{2}\left( \lambda_1 v^2+\lambda_2 v_{B-L}^2 - \frac{\lambda_1 v^2-\lambda_2 v_{B-L}^2}{\cos(2 \alpha )} \right) .
\end{align}
\end{subequations}
At the $\alpha \rightarrow 0$ limit, $h$ is reduced to the SM Higgs boson $\phi_H$.
For a small $\alpha \ll 1$, $h$ and $\phi$ are identified with the SM-like Higgs boson and the singlet-like scalar, respectively.
Thus, we take $m_h \simeq 125$ GeV.
The mixing angle $\alpha$ can be expressed in terms of $\lambda_3$ as
\begin{align}
\sin (2 \alpha ) & \simeq  \frac{2v v_{B-L}}{m_\phi^2-m_h^2}\lambda_3 = \frac{v m_{Z'}}{m_\phi^2 - m_h^2}\frac{\lambda_3}{g_{B-L}} ,
\end{align}
where we have used Eq.~(\ref{eq:Z'mass}) in the last equality.

In terms of $h$ and $\phi$, the scalar potential (\ref{eq:V}) is rewritten as
\begin{align}
 V 
 = & \frac{1 }{2}m_h^2h^2 +\frac{1}{2}m_{\phi}^2 \phi^2 +\frac{1}{6}C_{hhh}h^3 + \frac{1}{2}C_{hh\phi}h^2\phi + \frac{1}{2}C_{h\phi\phi}h\phi^2   + \frac{1}{6}C_{\phi\phi\phi}\phi^3 + \cdots ,
\end{align}
 where the ellipsis denotes quartic terms that are irrelevant for our following analysis.
The scalar trilinear couplings are given by
\begin{subequations}
\begin{align}
C_{hhh} &= 3 \frac{m_h^2 \left( v_{B-L} \cos^3\alpha- v \sin^3\alpha \right) }{v v_{B-L}}, \\
C_{hh\phi} &= \frac{\sin(2\alpha) (2 m_h^2+m_{\phi}^2)(v\sin\alpha+v_{B-L}\cos\alpha)}{2 v v_{B-L}}, \\
C_{h\phi\phi} &=\frac{\sin(2\alpha) (m_h^2+2 m_{\phi}^2)(v_{B-L}\sin\alpha-v\cos\alpha)}{2 v v_{B-L}}, \\
C_{\phi\phi\phi} &=3 \frac{ m_{\phi }^2 \left(v \cos^3\alpha +v_{B-L} \sin^3\alpha \right)}{v v_{B-L}},
\end{align}
\end{subequations}
where $C_{hh\phi}$ and $C_{h\phi\phi}$ are suppressed for the small mixing angle $\alpha$.

Similarly, each Yukawa coupling of the SM fermions $f$ and RH neutrinos $\nu_R$ with $h$ and $\phi$ is suppressed due to the Higgs mixing with the following factor 
\begin{subequations}
\begin{align}
& C_{hf} = \cos\alpha ,  \\
& C_{\phi f} = \sin\alpha, \\
& C_{h \nu_R} = -\sin\alpha,  \\
& C_{\phi \nu_R} = \cos\alpha, 
\end{align}
\end{subequations}
and the mixing suppression factors to the gauge couplings of the SM gauge bosons $V (W, Z, A)$ and $Z'$ to $h$ and $\phi$ are given by %
\begin{subequations}
\begin{align}
& C_{hV} = \cos\alpha,  \\
& C_{\phi V} = \sin\alpha, \\
& C_{h Z'} = -\sin\alpha,  \\
& C_{\phi Z'} = \cos\alpha .
\end{align}
\end{subequations}
From the above equations, we can read the interactions of $Z'$ and $\nu_R^i$ are affected by the Higgs mixing.
We will see later that the mixing is essential for the production of the sterile neutrino DM in some cases. 

\subsubsection{Decays of scalars and range of Higgs mixing}

Due to the Higgs mixing, decay rates of the SM-like Higgs boson into the SM particles are multiplied by $\cos^2\alpha$ in our model.
In addition, if the mass of the singletlike scalar $\phi$ is smaller than one half of the SM-like Higgs boson mass, another decay mode $h \rightarrow \phi\phi$ is possible.
Thus, the partial decay rates of $h$ are given by 
\begin{subequations}
\label{rate:h}
\begin{align}
\Gamma_h(h \rightarrow \mathrm{SM}) &=  \cos^2\alpha\Gamma_{h_\mathrm{SM}}(h_\mathrm{SM}\rightarrow \mathrm{SM}), \\
\Gamma_h(h \rightarrow \phi \phi) &= \frac{\sqrt{m_h^2-4 m_{\phi}^2}}{16\pi m_h^2} |C_{h\phi\phi}|^2 , \label{rate:h->phiphi} \\
\Gamma_h(h \rightarrow Z'Z') &=\frac{\sqrt{m_h^2-4 m_{Z'}^2}}{32\pi m_h^2} \sin^2\alpha\frac{m_h^4-4 m_h^2 m_{Z'}^2+12 m_{Z'}^4}{v_{B-L}^2} , \label{rate:h->Z'Z'}  \\
\Gamma_h(h \rightarrow NN) &=\sum_i \frac{1}{16 \pi m_h} \left( 1 -\frac{4 m_{N_i}^2}{m_h^2} \right)^{3/2} \sin^2\alpha\frac{ m_h^2 m_{N_i}^2 }{v_{B-L}^2} , \label{rate:h->NN} 
\end{align}
\end{subequations}
where $h_\mathrm{SM}\rightarrow \mathrm{SM}$  stands for the decay processes of SM Higgs boson into all final states in the SM model and its decay rate is $\Gamma_{h_\mathrm{SM}}(h_\mathrm{SM}\rightarrow \mathrm{SM}) \simeq 4 $ MeV. 
Here, $m_{N_i}$ is the mass of sterile neutrinos and $m_{N_i} \simeq m_{\nu_{R}^i}$ for small active-sterile mixings.

The total decay rate $\Gamma_h$ is given by the sum of Eqs.~(\ref{rate:h}).
For $m_{\phi} < m_h/2$, the current constraints on exotic decay of the SM Higgs boson as $\text{Br}(h \rightarrow \text{invisible}) \lesssim 19~\%$~\cite{ParticleDataGroup:2020ssz} can be recast as $\sin\alpha \lesssim 0.2$. For $m_{\phi} > m_h/2$, the obtained range $\alpha < \mathcal{O}(0.1)$ is also consistent with measurements in the LHC~\cite{ATLAS:2019nkf}.

In the mass spectra of our interest, the singletlike scalar $\phi$ decays dominantly into pairs of the SM fermions through
 the Higgs mixing with the rate
\begin{equation}
\label{rate:phi}
\Gamma_\phi(\phi \rightarrow \mathrm{SM}) =\sin^2\alpha\Gamma_{h_\mathrm{SM}}(\phi \rightarrow \mathrm{SM}),
\end{equation}
where $\Gamma_{h_\mathrm{SM}}(\phi \rightarrow \mathrm{SM})$ expresses the decay rate of $\phi$ with the same SM interactions of $h$, and the scale of running parameters is taken at $m_\phi$ in the calculation.
The following decay modes with partial decay rates
\begin{subequations}
\begin{align}
\Gamma_\phi(\phi \rightarrow hh) &= \frac{\sqrt{m_\phi^2-4 m_h^2}}{16\pi m_\phi^2}|C_{hh \phi}|^2 ,\\
\Gamma_\phi(\phi\rightarrow Z'Z') &=\frac{\sqrt{m_\phi^2-4 m_{Z'}^2}}{16\pi m_\phi^2} \cos^2\alpha\frac{m_\phi^4-4 m_\phi^2 m_{Z'}^2+12 m_{Z'}^4}{v_{B-L}^2} ,\\
\Gamma_\phi(\phi\rightarrow NN) &=\sum_i \frac{1}{16 \pi m_\phi} \left( 1 -\frac{4 m_{N_i}^2}{m_\phi^2} \right)^{3/2}\cos^2\alpha  \frac{m_\phi^2 m_{N_i}^2}{ v_{B-L}^2} ,
\end{align}
\end{subequations}
 can also open depending on the mass spectrum. 
These, however, are negligible compared with Eq.~(\ref{rate:phi}); $\Gamma_\phi(\phi \rightarrow hh) $ is due to $v_{B-L} \gg v$, and the others are also with the suppression by $v_{B-L}$ unless we take $\alpha \rightarrow 0$. 
Therefore, we find typically $\Gamma_{\phi} \sim \sin^2\alpha$ MeV.
Bounds on the Higgs mixing between a light scalar and the SM-like Higgs boson have been derived from the LEP experiments similarly~\cite{LEPWorkingGroupforHiggsbosonsearches:2003ing}, and the range $\alpha<\mathcal{O}(0.1)$ coincides with the bounds as well. 

If the singletlike scalar is lighter than about a few GeV, the constraints from meson decays by the LHCb~\cite{LHCb:2015nkv,LHCb:2016awg}
 and CHARM~\cite{CHARM:1985anb} are more stringent than ATLAS, CMS, and the LEP.
In such a mass range, the bound $10^{-5} \lesssim \sin\alpha \lesssim 10^{-4}$ has been obtained~\cite{Winkler:2018qyg}
 where the lower bound is set by demanding that the lifetime of $\phi$ must be shorter than $\mathcal{O}(0.1)$ seconds so as not to affect the BBN~\cite{Fradette:2017sdd}. 

\section{The Boltzmann equation}
\label{sec:bolt}

In this section we describe our Boltzmann equations to calculate the evolution and abundance of the sterile neutrino DM and $Z'$ via freeze-in production.
Here and hereafter the DM is denoted with $N$, which is a suitable one among the three sterile neutrinos.
The Boltzmann equations for the number density of $N$ and $Z'$ are given by
\begin{subequations}
\begin{align}
 \frac{d n_{N}}{dt} + 3 H n_{N} =& \sum_{i,j} \langle\sigma v(ij \rightarrow NN)\rangle (n_i n_j -n_N^2)
  + \sum_{i} \langle \Gamma(i \rightarrow NN)\rangle n_i , \label{Eq:BoltN_n} \\
 \frac{d n_{Z'}}{dt} + 3 H n_{Z'} =& \sum_{i,j} \langle\sigma v (ij\rightarrow Z' Z') \rangle( n_i n_j - n_{Z'}^2)+ \sum_i \langle\Gamma(i\rightarrow  Z'Z')\rangle  n_i  \nonumber \\
 &  + \sum_{i,j,k} \langle\sigma v ( Z' i\rightarrow jk)  n_i \rangle \left( n_{Z'}- n^{\mathrm{eq}}_{Z'} \right) 
 - \sum_{i,j} \langle\Gamma(Z'\rightarrow i j)\rangle  (n_{Z'}- n^{\mathrm{eq}}_{Z'}), \label{Eq:BoltX_n} 
\end{align}
 \label{eq:boltz-eq}
\end{subequations}
where $i, j$, and $k$ are possible initial and final states in reactions, and $n_i$ is the number density of $i$-particle.
Note that $N$ and $Z'$ productions from the decays of intermediate particles produced on-pole in the first term are treated properly to avoid double-counting in Eqs.~\eqref{eq:boltz-eq}.
The cosmic expansion rate $H$ in the radiation-dominated (RD) universe is given by
\begin{align}
 H^2 &\equiv  \left(\frac{\dot{a}}{a}\right)^2 = \frac{1}{3M_P^2}\rho_r,  \label{Eq:Fre} \\
 \rho_r &= \frac{\pi^2 g_*}{30}T^4,  \label{Eq:rhor} 
\end{align}
 where $a$ is the scale factor of the universe and dot denotes derivative with respect to the cosmic time $t$.
The reduced Planck mass is $M_P\simeq 2.4\times 10^{18}$ GeV, $\rho_r$ is the energy density of radiation with the temperature $T$ and $g_*$ denotes the number of relativistic degrees of freedom.

A thermally averaged product of the scattering cross section and the relative velocity in Eqs.~(\ref{Eq:BoltN_n}) and (\ref{Eq:BoltX_n}) are given by~\cite{Gondolo:1990dk}
\begin{align}
	\langle\sigma v\rangle n_i n_j & =
	\frac{T}{32\pi^4}\sum_{i, j}\int^{\infty}_{(m_i+m_j)^2} ds g_i g_j p_{ij} 4E_{i}E_j \sigma v K_1\left(\frac{\sqrt{s}}{T}\right) ,
\end{align}
 and
\begin{align}
4E_{i}E_j \sigma v 
& \equiv \prod_f \int \frac{d^3 p_f}{(2\pi)^3}\frac{1}{2E_f}\overline{|\mathcal{M}|^2}(2\pi)^4 \delta^{(4)}(p_i+p_j-\sum p_f) \nonumber \\
 & = \frac{1}{16 \pi}\frac{2|q_f|}{\sqrt{s}}\int \overline{|\mathcal{M}|^2} d\cos\theta , \\
2|q_f| & = \sqrt{s-4 m_N^2}, \\
 p_{ij} & \equiv \frac{\sqrt{s-(m_i+m_j)^2}\sqrt{s-(m_i-m_j)^2}}{2\sqrt{s}},
\end{align}
where $i$ is an initial state with mass $m_i$, energy $E_i$, and internal degrees of freedom $g_i$.
The center-of-mass energy squared is given by $s = (E_i + E_j)^2$ and three-momentum of a final-state particle is denoted by $q_f$.
$K_i(z)$ is the modified Bessel function of the $i$th kind.

We consider the vanishing limit of the active-sterile mixings unless otherwise stated in the following analyses.
This is because our purpose is to investigate the production of $N$ dominantly through the $g_{B-L}$ gauge interaction.
For the very small $g_{B-L}$ we are interested in, the first term in right-handed side of Eq.~(\ref{Eq:BoltX_n}) representing the pair productions $ij \rightarrow Z'Z'$ is actually negligible compared with the other terms. 
This is because the cross sections for such pair productions are suppressed by $g_{B-L}^4$ while processes described in other terms are suppressed by only $g_{B-L}^2$.

The second terms in the right-handed side of Eqs.~(\ref{Eq:BoltN_n}) and (\ref{Eq:BoltX_n}) represent the production by the decay of an $i$-particle, and 
\begin{align}
	\langle\Gamma(i \rightarrow NN \, \mathrm{or}\, Z'Z')\rangle =\frac{ K_1\left(\frac{m_i}{T}\right)}{K_2\left(\frac{m_i}{T}\right) } \Gamma(i \rightarrow NN \,\mathrm{or}\, Z'Z'),
\end{align}
is the thermal averaged partial decay rate of the $i$-particle which is suppressed for a high temperature, $T \gg m_i$, by the time dilation. 

The third term in right-handed side of Eq.~(\ref{Eq:BoltX_n}) principally denotes the processes of $f\bar{f}\leftrightarrow Z'\gamma$, $f\gamma\leftrightarrow f Z' $ and $\bar{f}\gamma\leftrightarrow \bar{f}Z'$.
The thermal averaging of $\sigma v n$ is defined as
\begin{align}
\langle \sigma v n_i \rangle n_{Z'}^{\mathrm{eq}}
& \equiv \frac{T}{32\pi^4}\int_{(m_i+m_{Z'})^2}^{\infty} ds g_i g_{Z'} p_{i{Z'}} (4E_i E_{Z'} \sigma v)K_1\left(\frac{\sqrt{s}}{T}\right) , 
\end{align}
with
\begin{align}
n_{Z'}^{\mathrm{eq}} &= \frac{T}{2\pi^2}g_{Z'}m_{Z'}^2K_2\left(\frac{m_{Z'}}{T}\right) ,
\end{align}
where the superscript ``$\mathrm{eq}$" stands for the equilibrium value.
Those turn out to be actually negligible compared with the fourth term. 
This can be understood from the fact that the cross sections of those $\gamma -Z'$ scatterings are suppressed by $g_{B-L}^2 \alpha_\mathrm{em}$
 with $\alpha_\mathrm{em}=e^2/(4\pi)$, while the following fourth term is suppressed by only $g_{B-L}^2$. 

The fourth term in Eqs.~(\ref{Eq:BoltX_n}) denotes the decay and inverse decay of $Z'$.
The extra neutral gauge boson $Z'$ decays into all fermions charged under the $U(1)_{B-L}$.
The partial decay widths of $Z'$ are given by
\begin{subequations}
\begin{align}
\Gamma_{Z'}(Z'\rightarrow f\bar{f}) &=\sum_f \frac{g_{B-L}^2 q_{B-L}^2 N_c}{12 \pi m_{Z'}^2} \left(2 m_f^2+m_{Z'}^2\right) \left(m_{Z'}^2-4 m_f^2 \right)^{1/2} , \\
\Gamma_{Z'}(Z'\rightarrow NN) &=\sum_{i} \frac{g_{B-L}^2}{24 \pi m_{Z'}^2} \left(m_{Z'}^2-4 m_{N_i}^2 \right)^{3/2} ,
\end{align}
\end{subequations}
 where the number of color $N_c=3$ is for quark final states. 
In this paper, we consider the situation where $h$ and $\phi$ are enough heavier than $Z'$. Then, the total decay rate is given by the sum of those.
The inverse decay is the most efficient process to thermalize $Z'$ disregarded in previous studies. This is our new observation in this work. 
Because of this efficient thermalization of $Z'$, we find the new freeze-in scenario
 where the $Z'Z'\rightarrow NN$ is the dominant production mode for $m_{Z'} \gtrsim 1$ MeV.
On the other hand, for $m_{Z'} \lesssim 1$ MeV, the magnitude of the $B-L$ gauge coupling $g_{B-L}$ turns out to be smaller than $\mathcal{O}(10^{-12})$ to avoid the overproduction of light $Z'$ as dark radiation.

\section{Abundance of sterile neutrino DM}
\label{sec:dm}
In this section we present the parameter region where cosmological constraints are satisfied and the observed DM abundance is explained with $N$. The following three mass spectra are considered.
Dominant processes of the DM production depend on the spectra, and hence the obtained regions are different in the three spectra. 
Before we show the results in detail, we briefly summarize those as
\begin{enumerate}
\item[A] {\bf Heavy gauge boson region $m_{Z'} > 2m_N$} : The DM $N$ and $Z'$ are not thermalized in this case. 
The $Z'$ are dominantly produced on shell by dominantly its inverse decay.
The DM is produced by the subsequent nonthermal decay $Z' \to NN$.
By taking the free-streaming length constraints into account for $m_{Z'} \lesssim 100$ GeV, the allowed mass range of $N$ turns out to be $m_N \gtrsim 1$ MeV and $g_{B-L} > 10^{-12}$. 
\item[B] {\bf Light gauge boson region $2m_N > m_{Z'} > 1$ MeV} : $Z'$ is thermalized by its inverse decay.
The DM is dominantly produced via pair annihilation $Z' Z' \to N N$. In this case, the DM abundance depends on the mass of $\phi$ due to the $s$-channel exchange of $\phi$. Taking $m_N = 1$ GeV and $1$ GeV $ < m_\phi < 100$ GeV, the allowed region is found in $10^{-10} < g_{B-L} < 10^{-6}$ and $10^{-3}$ GeV $< m_{Z'} < 2$ GeV.
\item[C] {\bf Very light gauge boson region $2m_N >  1$ MeV $> m_{Z'}$ } : 
The $B-L$ gauge coupling must be $g_{B-L}\lesssim 10^{-12}$ to avoid the BBN and the CMB constraints.
The DM must be dominantly produced from the scatterings of the SM particles and $\phi$ via $\phi/h$ $s$-channel exchange. 
\end{enumerate}

\subsection{Heavy gauge boson region $m_{Z'} > 2m_N$}
\label{sec:heavy}

The freeze-in DM production by the mediation of the extra gauge boson can be effective for $m_{Z'} > 2m_N$.
Under this mass spectrum, the production of $N$ can be dominated by the nonthermal decay of ${Z'}$.
Since not only $N$ but also ${Z'}$ cannot be thermalized, we need to simultaneously solve the Boltzmann Eqs.~(\ref{Eq:BoltN_n}) and (\ref{Eq:BoltX_n}), which are rewritten as
\begin{subequations}
\begin{align}
\left(\frac{dx}{dt}\right) \frac{d Y_{N}}{dx} =& \langle\Gamma(Z' \rightarrow NN)\rangle Y_{Z'}  , \label{Eq:BoltN_Y}  \\
 \left(\frac{dx}{dt}\right)\frac{d Y_{Z'}}{dx} =& \langle\Gamma(\phi\rightarrow Z'Z')\rangle Y_{\phi}-\sum\langle\sigma v (Z' i\leftrightarrow jk ) n_i\rangle  (Y_{Z'}-Y^{\mathrm{eq}}_{Z'}) \nonumber \\
 &  - \sum\langle\Gamma(Z'\leftrightarrow i j)\rangle (Y_{Z'} - Y^{\mathrm{eq}}_{Z'}),  \label{Eq:BoltX_Y} 
\end{align}
\end{subequations}
 where $x \equiv M/T$ with $M$ being a mass scale for the normalization is a dimensionless variable. 
The yield abundance $Y_i \equiv n_i/s$ is defined as the ratio of the number density to the entropy density
\begin{align}
 s = \frac{2\pi^2 g_{*S}}{45}T^3 ,
\end{align}
with $g_{*S}$ being the total relativistic degrees of freedom for the entropy.
Here, the pair production of $N$ by scattering $i j \rightarrow NN$ is dominated by the resonant processes of $s$-channel $Z'$ mediation from $f\bar{f}$ initial states. 
Since we have included the inverse decay of $Z'$, $f\bar{f}\rightarrow Z'$, and the decay of $Z'$ into $NN$, we have discarded the term for $f\bar{f} \rightarrow NN$ to avoid the double counting.

\begin{figure}[tbh]
\centering
\epsfig{file=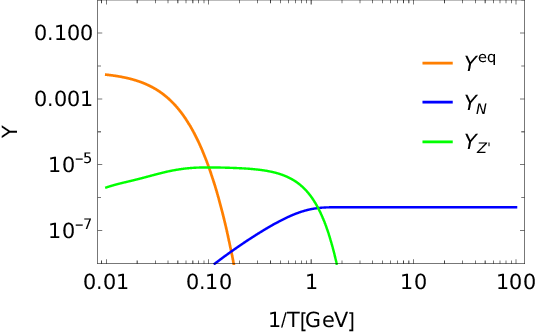, width=12cm,angle=0}
\caption{
Evolution of the yields of ${Z'}$ and $N$ for 
$m_{Z'} = 100$ GeV, $g_{B-L}=5 \times 10^{-10}$, $m_N=1$ MeV, $\alpha=0$.
 }
\label{fig:heavy_evol}
\end{figure}

We show, in Fig.~\ref{fig:heavy_evol}, the typical evolution of $Y_{Z'}$ and $Y_N$ for $m_{Z'} = 100$ GeV,
 $g_{B-L}=5 \times 10^{-10}$, $m_N=1$ MeV, and $\alpha=0$.\footnote{This condition $\alpha=0$ is taken to suppress scalar mediated processes and is not necessarily satisfied exactly. However, when $\alpha$ is sufficiently large, the scalar mediated processes easily dominate over the $Z'$ mediated processes and it is reduced to the Higgs portal freeze-in Majorana DM model.}
The orange curve is the thermal equilibrium yield value of the $Z'$ boson.
The blue and green solid curves represent values of $Y_{Z'}$ and $Y_N$, respectively.
Here, we have confirmed that the $\gamma - Z'$ scatterings are negligible compared to the inverse decay of $Z'$, as mentioned above.
Once $Y_N(x \rightarrow \infty)$ is obtained, the present relic density is evaluated as
\begin{align}
\Omega h^2  = \frac{m_N}{\rho_\mathrm{crit}/s_0}Y_N
\end{align}
where $(\rho_\mathrm{crit}/s_0)^{-1} = 2.8 \times 10^8/\mathrm{GeV}$ is given by the present entropy density $s_0$ and the critical density $\rho_\mathrm{crit}=3 M_P^2 H_0^2$ with $H_0$ being the present Hubble parameter.

Since the DM $N$ are produced by the decay of out-of-equilibrium $Z'$, under the mass spectrum of $m_{Z'} \gg m_N$, $N$ could be generated with a large momentum of about a half of the $Z'$ mass.  
Such an energetic $N$ can have a large free-streaming length and erase small scale structures.
The resultant comoving free-streaming scale can be calculated as~\cite{Kolb:1990vq} 
\begin{align}
R_f &= \int^{t_{\mathrm{mre}}}_{t_{\mathrm{dec}}}\frac{v(t')}{a(t')}dt' \nonumber \\
 &\simeq
\frac{1}{2 a_{\mathrm{mre}} H_{\mathrm{mre}}} \frac{a_{\mathrm{nr}}}{a_{\mathrm{mre}}} \left(
\log \left(1+\frac{1}{\sqrt{1+\left(\frac{a_{\mathrm{nr}}}{a_{\mathrm{mre}}}\right)^2}} \right)
- \log \left(1-\frac{1}{\sqrt{1+\left(\frac{a_{\mathrm{nr}}}{a_{\mathrm{mre}}}\right)^2}}\right)\right),
\label{Eq:Rf}
\end{align}
 with the velocity $v$. The three-momentum of produced DM normalized by the mass
\begin{align} 
\frac{p_N}{m_N} = u = \frac{v}{\sqrt{1-v^2}} ,
\end{align}
 whose initial value 
\begin{align} 
 u(t_{\mathrm{dec}})= \frac{\sqrt{m_{Z'}^2-4m_N^2}}{2 m_N} ,\label{Eq:iniu}
\end{align}
 is given at the time of the $Z'$ decay , $t_{\mathrm{dec}} = 1/\Gamma_{Z'}$, red-shifts inversely proportional to the scale factor as $u \propto a(t_{\mathrm{dec}})/a(t) = a_{\mathrm{dec}}/a(t)$. 
The scale factor $a_{\mathrm{nr}}$ is one at the time $t_{\mathrm{nr}}$ when $N$ becomes nonrelativistic, {\it i.e.}, $u=1$, which is evaluated as
\begin{align} 
u(t_{\mathrm{dec}}) \frac{a_{\mathrm{dec}}}{a_{\mathrm{nr}}} =1 . \label{def:anr}
\end{align}
At the time of the matter-radiation equality, $t_{\mathrm{mre}}$, 
\begin{align} 
\frac{a_0}{a_\mathrm{mre}} & = 1+ z_{\mathrm{mre}} = \frac{\Omega_m h^2}{\Omega_r h^2}, \label{Eq:zeq} \\
H_{\mathrm{mre}} &= \sqrt{2 \Omega_m } H_0 (1+ z_{\mathrm{mre}})^{3/2},  \label{Eq:Heq}
\end{align}
where $\Omega_{r(m)}$ is the density parameter of the radiation(matter), and $a_0$ and $H_0$ are the present scale factor and Hubble parameter, respectively.
We find 
\begin{align} 
\frac{a_{\mathrm{nr}}}{a_{\mathrm{mre}}} =\frac{a_{\mathrm{dec}}}{a_{\mathrm{mre}}} u(t_{\mathrm{dec}}) =  \sqrt{\frac{t_{\mathrm{mre}}}{t_{\mathrm{dec}}}} u(t_{\mathrm{dec}}) , \label{Eq:anraeq-ratio}
\end{align}
from Eq.~(\ref{def:anr}) and $a \propto \sqrt{t}$ in the RD universe. 
By substituting Eqs.~(\ref{Eq:iniu}), (\ref{Eq:zeq}), (\ref{Eq:Heq}), and (\ref{Eq:anraeq-ratio}) into Eq.~(\ref{Eq:Rf}), we can evaluate the free-streaming length as $\lambda_{\mathrm{fs}} = a_0 R_f$.
While the non-negligible free-streaming length would be interesting from the viewpoint of small-scale problems in cold dark matter model, it should not be larger than sub-Mpc.
The recent constraints on warm DM or the free-streaming scale have been reported in Ref.~\cite{Irsic:2017ixq,DES:2020fxi}.

\begin{figure}[tbh]
\centering
\epsfig{file=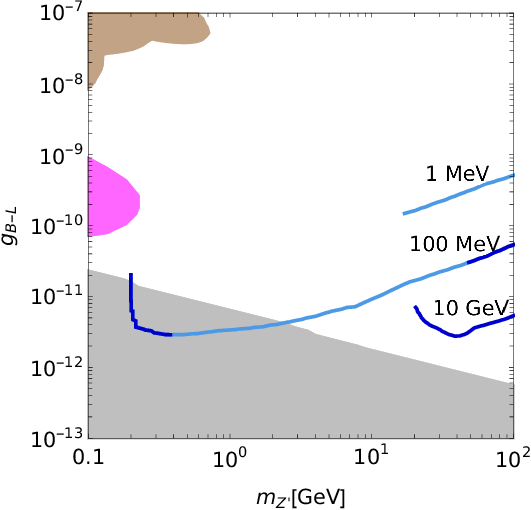,width=12cm,angle=0}
\caption{
Contours of $\Omega h^2=0.12$ for several values of $m_N$ with bluish lines. The region with $1/\Gamma_{Z'} > 0.1$ second is shaded by gray, and regions excluded by the SN1987A constraint and beam dump experiments are colored magenta and brown, respectively.
 }
\label{fig:heavy_results}
\end{figure}

The contours of the observed DM density $\Omega h^2=0.12$ in Fig.~\ref{fig:heavy_results} appear as blueish curves for $m_N = 1$ MeV, $100$ MeV, and $10$ GeV, from top to bottom, respectively.
The dark and light blue curves correspond to $\lambda_{\mathrm{fs}} < 0.01$ Mpc, and $0.01$ Mpc $< \lambda_{\mathrm{fs}} < 0.1$ Mpc, respectively. 
The left endpoint of the light blue curve for $M_N=1~\mathrm{MeV}$ corresponds to $\lambda_{\mathrm{fs}} = 0.1$ Mpc.
From the line for $M_N=100~\mathrm{MeV}$, we can read that the free-streaming length is shortened not only for $m_{Z'} \gtrsim 2 m_N$ but also for $m_{Z'} \gg m_N$. In the latter case, the $Z'$ decay happens relatively early and thus there is enough time to be redshifted for the momentum of $N$.
As long as we consider $m_{Z'} \lesssim 100 $ GeV, the mass of the DM produced by this mechanism must be larger than about MeV to have a small enough free-streaming length.
The region where the lifetime of the $Z'$ boson is longer than $0.1$ second is shaded in gray.
It is ruled out because the decay of such long-lived abundant $Z'$ bosons produces energetic particles and destroys the light elements synthesized by big bang nucleosynthesis.
We employed the SN1987A constraint from Ref.~\cite{Croon:2020lrf} as a reference and the excluded region is colored magenta. For the constraint see also recent other discussions~\cite{Shin:2021bvz,Caputo:2022rca}.
The region colored brown is excluded by electron and proton beam dump experiments (\cite{Feng:2022inv,Asai:2022zxw} and references therein).\footnote{The excluded regions by beam dump experiments and SN1987A have been derived under the assumption that $Z'$ does not decay into RH neutrinos. Therefore, although Fig.~\ref{fig:heavy_results} is presented for the $m_{Z'} > 2m_N$ case, the shaded area would not be exact. We, however, show them for reference purposes, because the excluded region is very far away from the parameter region of our interest and the viable parameter space of DM is unaffected. Since sterile neutrino $N$ is also a very weakly-interacting light particle, $N$ in addition to $Z'$ would contribute the energy loss of supernovae. However, while the production cross section of $Z'$ is as $\sigma \propto g_{B-L}^2$, that of $N$ pair is $\sigma \propto g_{B-L}^8$ because the vertex of $f \rightarrow f N N$ is induced by the one loop diagram running $N$ and $Z'$ with $f$ being SM particles. Thus, the latter is negligible compared with the former, for $g_{B-L} \ll 1$.
}

\subsection{Light gauge boson region : $ 2m_N > m_{Z'} > 1$ MeV }
\label{sec:light}

Next, we consider the mass spectrum of $2m_N > m_{Z'} > 1$ MeV, in which the lower limit corresponds to the typical temperature of BBN.
If we consider a smaller mass of $Z'$ or larger coupling $g_{B-L}$ than the parameter sets studied in Sec.~\ref{sec:heavy}, the $Z'$ gauge boson is fully thermalized by its decay and inverse decay described by
 the fourth terms in the right-handed side of Eq.~(\ref{Eq:BoltX_n}).

\begin{figure}[!tbh]
	\begin{center}
		\begin{tabular}{cc}
\epsfig{file=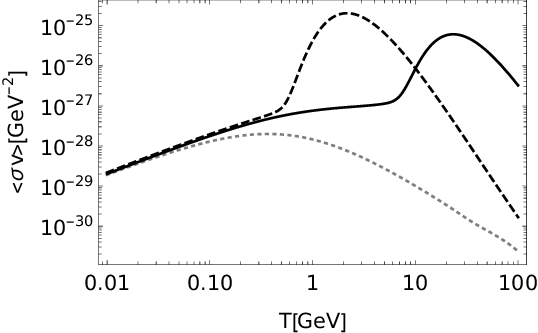, width=8cm,angle=0}
			 &
\epsfig{file=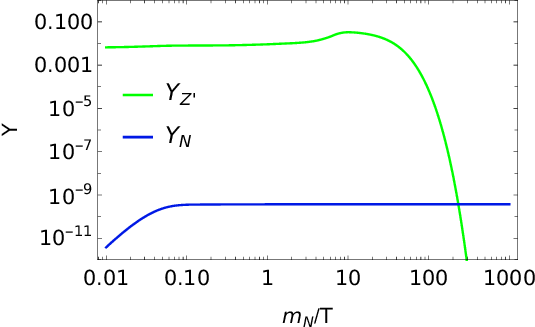, width=8cm,angle=0}
		\end{tabular}
	\end{center}
	\caption{Plots are made for $\alpha=0$, $m_{Z'} = 0.1$ GeV, $g_{B-L}=2 \times 10^{-8}$, $m_N=1$ GeV.
\textit{Left}: $\langle \sigma v\rangle$ for $m_{\phi} = 1$ GeV (gray dotted), $10$ GeV (black dashed), and $100$ GeV (black solid). 
 \textit{Right}:
Evolution of the yields of $Z'$ (green curve) and $N$ (blue curve) for $m_{\phi} = 100$ GeV. For this parameter set, $\Omega_N h^2 \simeq 0.1$ is reproduced.}
\label{fig:light_evol}
\end{figure}

As a result, $N$ is dominantly produced by the pair-production processes $Z'Z' \rightarrow NN$ via $t$($u$)-channel $N$ exchange processes and $s$-channel $\phi$ and $h$ exchange processes.
The remarkable feature is the $m_{\phi}$ dependence of the DM abundance.
For $m_{Z'} < m_N$, the scattering cross section of this process grows with respect to $s$ at lower energy than the mediator mass scale due to the longitudinal mode of $Z'$. 
In fact, the leading part of invariant amplitude squared, whose full expression is noted in Appendix, at a large $s > m_N^2 \gg m_{Z'}^2 $ is
\begin{align}
\int \sum |\mathcal{M}|^2 d\cos\theta 
 \sim 
\frac{32 g_{B-L}^4}{m_{Z'}^4} \left(\frac{2 m_N^4 s^2}{m_N^2 \left(s-4 m_{Z'}^2\right)+m_{Z'}^4}-\frac{2 m_N^2 s^3}{\Gamma_{\phi}^2 m_{\phi}^2+\left(m_{\phi}^2-s\right)^2}\right),
\end{align}
 where $\theta$ is the scattering angle.
The first term from the $t(u)$-channel $N$ exchange processes grows linearly with respect to the center-of-mass energy $s$, as $s$ becomes large. 
The second term from $s$-channel $\phi$ exchange processes becomes comparable for $s \gtrsim m_{\phi}^2$ and cancels with the first term at $s \gg m_{\phi}^2$. 
The $m_{\phi}$ dependence in the thermally averaged cross section for $m_N=1$ GeV is shown in the left panel of Fig.~\ref{fig:light_evol}. 
Each curve is for $m_{\phi} = 1$ GeV (gray dotted), $10$ GeV (black dashed), and $100$ GeV (black solid). 
The black dashed and solid curves for $m_{\phi} > m_N$ have a sharp peak at $T \sim m_{\phi}/5$ due to the resonance pole of $\phi$ and decrease at the high temperature region $T > \mathcal{O}(m_{\phi})$.
It should be emphasized here that the DM production takes place most effectively at not $T \sim m_N$ but $T \sim m_{\phi}/5$, unlike most freeze-in scenarios with renormalizable couplings. 
Hence, the result is sensitive to the mass of mediator $m_{\phi}$.

This characteristic dependence of $m_\phi$ can be seen concretely in an example of the evolution of $Y_{Z'}$ and $Y_N$ for
$m_N=1$ GeV, $m_\phi=100$ GeV, $m_{Z'} = 0.1$ GeV, $g_{B-L}=2 \times 10^{-8}$, $\alpha=0$ presented in right panel of Fig.~\ref{fig:light_evol}.
The blue and green curves represent values of $Y_N$ and $Y_{Z'}$, respectively.
The $Z'$ boson is thermalized and its yield follows the thermal value.
The slight increase of $Y_{Z'}$ around $m_N/T \sim 10$ is due to the change of $g_*$ at the quark hadron transition. We can see that $N$ is gradually generated until the temperature becomes as low as $\sim m_{\phi}/5$ not the DM mass $m_N$. The energy density of $Z'$ decreases due to the Boltzmann suppression. Thus, the energy density of $Z'$ gets negligible by the onset of the BBN. 

\begin{figure}[!htb]
\centering
\epsfig{file=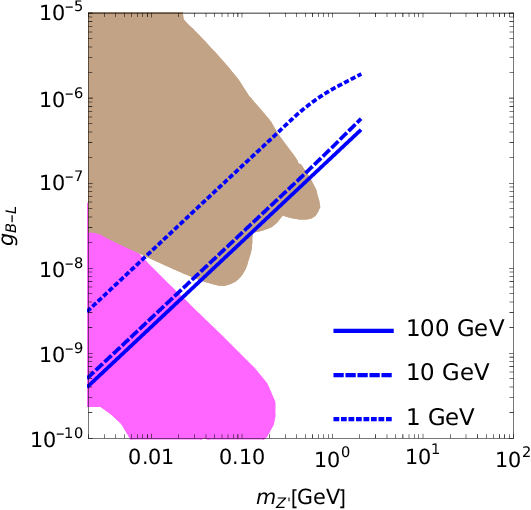,width=12cm,angle=0}
\caption{
Same as Fig.~\ref{fig:heavy_results}, but for $ 2m_N > m_{Z'} > 1$ MeV. 
Contours of $\Omega h^2=0.12$ with $m_N=1$ GeV are shown by blue dotted, dashed, and solid curves for $m_{\phi} =1, 10, 100$ GeV, respectively.
 }
\label{fig:light_results}
\end{figure}

The contours of $\Omega h^2=0.12$ in this mass spectrum are shown in Fig.~\ref{fig:light_results}.
The colored, excluded regions are the same as Fig.~\ref{fig:heavy_results}.
The curves shift to the right or left when we increase or decrease $m_N$. 
We also note that parameters to reproduce the observed DM abundance interestingly lie in the reach of future experiments for long-lived particle search or beam dump experiments.

%

\subsection{Very light gauge boson region : $ 2m_N > 1$ MeV $ > m_{Z'}$}
\label{sec:verylight}

Finally, we consider the mass spectrum of $ 2m_N > 1$ MeV $ > m_{Z'}$.
If such a light $Z'$ boson is thermalized by its inverse decay, the contribution of $Z'$ to the energy density of relativistic degrees of freedom at the BBN conflicts with the observation.
To avoid this, the gauge coupling constant must be much smaller than about $10^{-10}$.
The smallness of the coupling makes all processes for sterile neutrino production via $g_{B-L}$ gauge interaction negligible.
However, $f\bar{f} \rightarrow NN$, $W^+W^- (ZZ) \rightarrow NN$ and $hh ( h\phi, \phi\phi) \rightarrow NN$ of all the s-channel scalar ($h$ or $\phi$) exchanges are relevant processes to produce the DM $N$. 
Those processes get effective at the electroweak and $B-L$ broken vacuum, and thus only $b\bar{b}$ and $\tau\bar{\tau}$ initial states are dominant, $\phi\phi$ initial state could be non-negligible for some parameter sets, and the other initial states are negligible. 

\begin{figure}[htb]
\centering
\epsfig{file=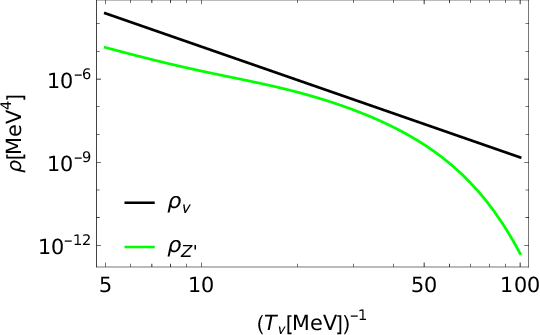,width=12cm,angle=0}
\caption{
The evolution of the energy density of $Z'$ (green) and one generation of neutrino (black) for $g_{B-L} = 5\times 10^{-12}$, $m_{Z'}=0.3$ MeV, $m_N = 1$ GeV, and $m_{\phi}= 500$ GeV.
 }
\label{fig:verylight_evolution}
\end{figure}

Since the evolution of abundance of $N$ and $Z'$ are independent in this case, two Boltzmann equations~(\ref{Eq:BoltN_n}) and (\ref{Eq:BoltX_n}) are decoupled. 
We can solve both individually.
First, let us see the evolution of the abundance of the gauge boson $Z'$,
\begin{align}
 \frac{d n_{Z'}}{dt} + 3 H n_{Z'} =& 
 - \sum\langle\Gamma(Z'\leftrightarrow i j)\rangle (n_{Z'}- n^{\mathrm{eq}}_{Z'})  ,
\end{align}
where we omit negligible inverse processes and negligible terms $\sigma v (ij\leftrightarrow Z' Z')$, $\Gamma(i\leftrightarrow  Z'Z')$, and the $\gamma -Z'$ scattering. 
The term in the right-handed side is the decay and the inverse decay of $Z'$. 
The final abundance is determined by $g_{B-L}$, because the magnitude of the rate is proportional to $g_{B-L}^2$.
In Fig.~\ref{fig:verylight_evolution}, the evolution of the energy density of $Z'$ for $g_{B-L} = 5\times 10^{-12}$ is shown and compared with that of one generation of neutrino. 
Because of the very long lifetime of $Z'$, the constraint from CMB is much more stringent than that from the BBN.
As seen in Fig.~\ref{fig:verylight_evolution}, even if the abundance of $Z'$ is sufficiently small at the time of the BBN $T_{\nu} \sim 1$ MeV, the energy fraction starts to increase for $T \lesssim m_{Z'}$ because the energy density decreases as $a^{-3}$ after $Z'$ becomes nonrelativistic. 
Thus, the extra radiation generated by $Z'$ decay could be significant at the recombination epoch and affects the temperature anisotropy of CMB unless the energy density of $Z'$ is small sufficiently at the BBN era.

While the evolution of the energy density of $Z'$ can be followed as above, the abundance of $Z'$ can be easily and directly estimated by the following single integration expression 
\begin{align}
Y_{Z'}  = & \int_{T_0}^{T_R} \frac{\sum\langle\Gamma(Z'\leftrightarrow i j)\rangle n_i n_j }{s T H}dT \nonumber \\
  = & \frac{135 \sqrt{10} M_P}{2 \pi^3} \int_{T_0}^{T_R} \frac{dT}{g_{*S} \sqrt{g_* } T^6}
 \sum_{i, j} \langle\Gamma(Z'\leftrightarrow i j)\rangle n_{Z'}^{\mathrm{eq}} ,
\end{align}
where $T_0$ is a low temperature before the decay of $Z'$. 
The $Z'$ decays into LH neutrino pairs at $t=1/\Gamma_{Z'}$.   
By comparing the energy density of neutrino of one species 
\begin{equation}
\rho_{\nu} = g_{\nu}\frac{7}{8}\frac{\pi^2}{30}T_{\nu}^4 ,
\end{equation}
where $g_{\nu}=2$ is the internal degrees of freedom of neutrinos and $T_{\nu}$ is the temperature of neutrinos, the energy density of decay products from $Z'$ can be parametrized as
\begin{align}
\Delta N_{\mathrm{eff}} \equiv \left.\frac{\rho_{Z'}}{\rho_{\nu}}\right|_{t=1/\Gamma_{Z'}}.
\end{align}

Similarly, by integrating Eq.~(\ref{Eq:BoltN_n}) from a low temperature $T_0$ to the reheating temperature after inflation $T_R$ as
\begin{align}
Y_N & = \int_{T_0}^{T_R} \frac{\langle\sigma v (i j \rightarrow NN)\rangle n_i n_j }{s T H}dT \nonumber \\
 & = \frac{135 \sqrt{10} M_P}{64 \pi^7} \int_{T_0}^{T_R} \frac{dT}{g_{*S} \sqrt{g_* } T^5}\sum_{i, j}\int^{\infty}_{(m_i+m_j)^2} ds g_i g_j p_{ij} 4E_{i}E_j \sigma v K_1\left(\frac{\sqrt{s}}{T}\right) ,
\end{align}
in the second equality, we have used Eqs.~(\ref{Eq:Fre}) and (\ref{Eq:rhor}).
In this framework, there are five free parameters associated with the production of DM $N$; $g_{B-L}, m_{Z'}, m_N, m_{\phi}$, and $\alpha$.
Although $\Omega h^2$ seems to be dependent on all the five parameters, it practically depends on only three of $g_{B-L} \sin(2\alpha)/m_{Z'}, m_{\phi}$, and $m_N$. 
We can find a simple scaling of the resultant abundance as
\begin{align}
\Omega h^2 & \propto \left(\frac{g_{B-L} \sin(2\alpha)}{m_{Z'}}\right)^2, 
\end{align}
because the coupling vertex appears only in this combination for the Higgs portal main processes.
As we discussed in Sec.~\ref{sec:model}, the Higgs mixing is constrained as $\alpha \lesssim 0.1$ for $m_{\phi} \gtrsim 10$ GeV and $\alpha \lesssim 10^{-4}$ for $m_{\phi} \lesssim$ several GeV.

\begin{figure}[!tbh]
	\begin{center}
		\begin{tabular}{cc}
\epsfig{file=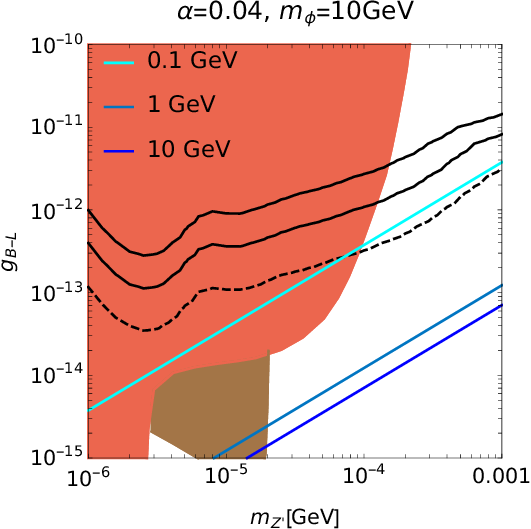, width=8cm,angle=0}
			 &
\epsfig{file=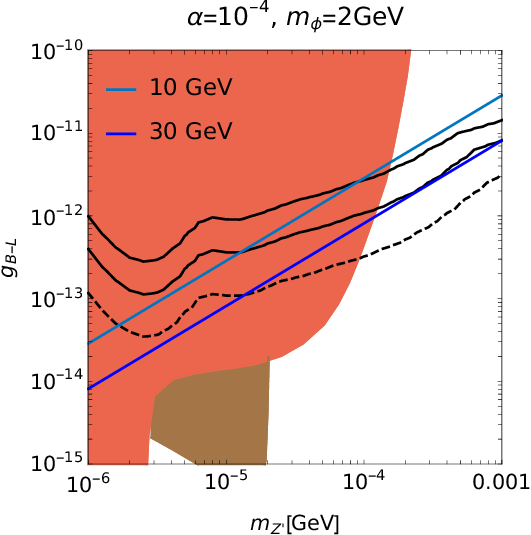, width=8cm,angle=0}
		\end{tabular}
	\end{center}
	\caption{
Contours of $\Omega h^2=0.12$ and $\Delta N_{\mathrm{eff}}$ presented with the parameter region excluded by the constraints 
from the horizontal branch star and red giant stars.
The contours of $\Omega h^2=0.12$ are drawn with bluish lines for some $m_N$ in key.
The contours with black (solid, solid and dashed) curves are
 for $\Delta N_{\mathrm{eff}} = 0.5, 0.2,$ and $ 0.06$ from top to bottom, respectively.
\textit{Left}: Contours for $\alpha=0.04$, and  $m_{\phi} = 10$ GeV.
\textit{Right}: Contours for $\alpha=10^{-4}$, and $m_{\phi} = 2$ GeV.
}
\label{fig:verylight_results}
\end{figure}

In Fig.~\ref{fig:verylight_results} the contours of $\Omega h^2=0.12$ are shown with bluish curves for some $m_N$ and for two sets of $\alpha$ and $m_{\phi}$.
Two black solid and one black dashed curves are contours of $\Delta N_{\mathrm{eff}}= 0.5, 0.2$, and $0.06$ from top to bottom, respectively. 
We note that extra relativistic degrees of freedom with $ 0.2 \lesssim \Delta N_{\mathrm{eff}} \lesssim 0.5$ would be favored to relax the so-called Hubble tension~\cite{Planck:2018vyg,Seto:2021xua,Seto:2021tad} and that $\Delta N_{\mathrm{eff}}= 0.06$ is the expected reach of the future experiment CMB-S4~\cite{Abazajian:2019eic}.
The excluded region by the horizontal branch star and the red giant star constraints are shaded with orange and brown, respectively~\cite{Redondo:2013lna,Heeck:2014zfa}. 
In the left panel, three lines of $\Omega_N h^2\simeq 0.12$ are shown for $m_N=0.1,1$, and $10$ GeV. 
The uneven intervals are due to the change of available modes. 
Namely, for a smaller $m_N$, the pair-production mode with heavy SM fermion initial states are suppressed at $T \sim m_N$.
The right plot is an example with smaller $m_{\phi}$ and $\alpha$. 
In this case, the larger $m_N > \mathcal{O}(10)$ GeV is required to reproduce the desired DM abundance, because the production cross section is strongly suppressed by the tiny mixing $\alpha$.

\section{Implication of sterile neutrino DM detection }
\label{sec:detection}

We have considered sterile neutrino DM whose mass is larger than MeV. 
The visible decay modes include $N \rightarrow \nu \gamma$, three body decay $N \rightarrow \nu f \bar{f}$ through off-shell $W^\pm$ and $Z$, and even hadronic mode.\footnote{For formulas of those decay rates, see for instance Refs.~\cite{Atre:2009rg,Ballett:2016opr}.}
While the usual keV scale sterile neutrino DM is searched for with x-ray lines induced by its radiative decay~\cite{Pal:1981rm}, the DM argued in this paper can be also probed by seeking other modes, such as the decay into $e^- e^+$ and the continuous spectrum of gamma rays from hadrons.

All the decays are induced by the SM processes through the active-sterile mixing for $m_N < m_{Z'}$.
On the other hand, for $m_N > m_{Z'}$, a new decay mode $N \rightarrow \nu Z'$ is additionally possible, and thus signals to search for $N$ are decided by the decay modes of $Z'$ depending on $m_{Z'}$.
In the very light gauge boson region, $Z'$ can decay into only neutrinos, while decaying modes into other fermions opens as the mass increases.

Therefore, some of the decay rates depend on whether $Z'$ is heavier than $N$ or not, for example, the rate of $N \rightarrow e^- e^+ \nu$.
Since $N \rightarrow \nu Z'$ is a tree-level process, the constraint on the active-sterile mixing may be more stringent than that from x-ray observations, especially for $m_N > m_{Z'}$ cases. 
The detailed study is beyond the scope of this paper, and we will evaluate this issue elsewhere.

\section{Conclusion}
\label{sec:concl}

We have evaluated the freeze-in production of $U(1)_{B-L}$ gauge interacting sterile neutrino DM by taking into account processes overlooked in the literature; principally the inverse decay of $Z'$ and the longitudinal mode effect in $Z'Z' \rightarrow NN$ scattering.
We have found that the inverse decay of $Z'$ indeed gives a non-negligible contribution to the production of $Z'$.

For $m_{Z'}> 2 m_N$ cases, the final value of $\Omega_N h^2$ agrees with the previous estimation.
Our finding for this mass spectrum case is that the constraint from the free-streaming length is more stringent than that has been thought. 
As the result, the mass of sterile neutrino DM under this mass spectrum must be larger than about one MeV as long as we assume $m_{Z'} \lesssim 100$ GeV.

For the other spectrum $m_{Z'} < 2 m_N$, the gauge coupling $g_{B-L} = \mathcal{O}(10^{-6})$ independent from the $m_{Z'}$ has been regarded as the viable parameter region to reproduce the desired DM abundance. 
We, however, have found that this is not correct.
Since $Z'$ is lighter than $N$, when $N$ is produced by $Z'Z' \rightarrow NN$, this cross section is enhanced by small $m_{Z'}^2$ and increases with respect to the energy until $s\sim m_{\phi}^2$, due to the longitudinal mode of $Z'$.
Thus, the resultant DM abundance depends on $m_{Z'}$ and $m_{\phi}$. 
In addition, for such a large coupling of $g_{B-L} \sim 10^{-6}$, $Z'$ can be thermalized and gives $\Delta N_\mathrm{eff} \sim 1 $ at the BBN epoch for $m_{Z'} \lesssim 1$ MeV. 
Moreover, CMB gives a more stringent constraint on the gauge coupling as $g_{B-L} \lesssim 10^{-12}$ than BBN, because the energy density of $Z'$ decreases slower than the background radiation after it becomes nonrelativistic.
Thus, for $ 2m_N > 1$ MeV $ > m_{Z'}$, the freeze-in production by the $Z'$ mediation is not available and the DM in this parameter region can be produced only by scalar portal scatterings.

Having the parameter space of consistent sterile neutrino DM mentioned above, the viable parameters for $m_{Z'}> 2 m_N$ lie far beyond the reach of the near future experiments of long-lived particle searches as shown in Fig.~\ref{fig:heavy_results}, while a large interesting parameter space in the spectrum of $1$ MeV $<m_{Z'} < 2 m_N$ is already constrained partially by the current experimental limits and will be probed more by the experiments in future. 
The case of the spectrum with $m_{Z'}<1 $MeV $ < 2 m_N$ will be examined by future measurements of $N_\mathrm{eff}$.

\section*{Acknowledgments}
We thank Masahiro Ibe for valuable comments on the IR divergence of the $Z'-\gamma$ scattering.
This work is supported, in part, by JSPS KAKENHI Grants No.~JP19K03860 and No.~JP19K03865 and MEXT KAKENHI Grant No.~21H00060 (O.~S.), 
and JSPS KAKENHI Grants No.~18K03651, No.~18H01210, No.~22K03622 and MEXT KAKENHI Grant No.~18H05543 (T.~S.).

\appendix

\section{Amplitude of the sterile neutrino pair-production processes}

%
We give explicit formulas of the invariant amplitude squared.

\subsection{$ f(p_1)\bar{f}(p_2) \rightarrow N(q_1) N(q_2)$}

%
\begin{align}
\overline{|\mathcal{M}|^2}d\cos\theta = 
\frac{m_f^2 m_N^2 N_c \sin^2(2\alpha) \left(s-4 m_f^2\right) \left(s-4 m_N^2\right) }{2 v^2 v_{B-L}^2}
\left| \frac{-1}{s-m_h^2-i \Gamma_h m_h}+\frac{1}{s-m_{\phi}^2 -i m_{\phi}\Gamma_{\phi}}\right|^2 ,
\end{align}
\begin{align}
4E_{i}E_j \sigma v (i j \rightarrow ) 
 = \frac{1}{16\pi}\sqrt{\frac{s-4m_N^2}{s}} \overline{|\mathcal{M}|^2}d\cos\theta , \\
p_{ij} = \frac{\sqrt{s-4m_f^2}}{2}.
\end{align}
%

\subsection{$ W(p_1)W(p_2) \rightarrow N(q_1) N(q_2)$}

%
\begin{align}
\overline{|\mathcal{M}|^2}d\cos\theta = &
\frac{g_2^2 m_N^2 \sin^2(2\alpha)  \left(s-4 m_N^2\right) \left(-4 m_N^2 \left(2 m_W^2+s\right)+4 m_N^4+16 m_W^4+s^2\right) }{36 m_W^2 v_{B-L}^2 } \nonumber \\
 & \times \left| \frac{-1}{s-m_h^2-i \Gamma_h m_h}+\frac{1}{s-m_{\phi}^2 -i m_{\phi}\Gamma_{\phi}}\right|^2 ,
\end{align}
\begin{align}
4E_{i}E_j \sigma v (i j \rightarrow ) 
 = \frac{1}{16 \pi}\sqrt{\frac{s-4 m_N^2}{s}} \int \overline{|\mathcal{M}|^2} d\cos\theta , \\
p_{ij} = \frac{\sqrt{s-4m_W^2}}{2}.
\end{align}
%

\subsection{$ Z(p_1)Z(p_2) \rightarrow N(q_1) N(q_2)$}

%
\begin{align}
\overline{ |\mathcal{M}|^2} d\cos\theta  = &
\frac{g_2^2 m_N^2 \sin^2(2\alpha) \left(s-4 m_N^2\right) \left(-4 m_N^2 \left(2 m_Z^2+s\right)+4 m_N^4+16 m_Z^4+s^2\right) }{36 m_Z^2 c_W^2 v_{B-L}^2 } \nonumber \\
 & \times 
\left| \frac{-1}{s-m_h^2-i \Gamma_h m_h}+\frac{1}{s-m_{\phi}^2 -i m_{\phi}\Gamma_{\phi}}\right|^2 ,
\end{align}
\begin{align}
4E_{i}E_j \sigma v (i j \rightarrow ) 
 = \frac{1}{16 \pi}\sqrt{\frac{s-4 m_N^2}{s}} \int \overline{|\mathcal{M}|^2} d\cos\theta  ,\\
p_{ij} = \frac{\sqrt{s-4m_Z^2}}{2}.
\end{align}
%

\subsection{$ (h, \phi)(p_1) (h, \phi)(p_2) \rightarrow N(p_1) N(p_2) $}

%
\begin{align}
i \mathcal{M} =& -\frac{ i m_N\bar{u}(p_1,m_N)v(p_2,m_N)}{v_{B-L}}\left( 
\frac{C_{h}\sin\alpha}{ (q_1+q_2)^2-m_h^2+i\Gamma_h m_h}
+\frac{C_{\phi}\cos\alpha}{ (q_1+q_2)^2-m_{\phi}^2+i m_{\phi}\Gamma_{\phi}}
 \right)  , 
\end{align}
where we have omitted $t(u)$-channel $N$ exchange contributions because those are suppressed by $m_N^2/v_{B-L}^2$ and are thus negligible.

\begin{align}
\overline{  |\mathcal{M}_s|^2} d\cos\theta = 
\frac{4 m_N^2 (s-4 m_N^2)}{v_{B-L}^2}
\left| \frac{C_{h}\sin\alpha}{ s-m_h^2+i\Gamma_h m_h}
+\frac{C_{\phi}\cos\alpha}{s-m_{\phi}^2+i m_{\phi}\Gamma_{\phi}}\right|^2 ,
\end{align}
\begin{align}
(C_{h}, C_{\phi}) = 
&  (C_{h\phi\phi}, C_{\phi\phi\phi} )   \quad \mathrm{for}   \quad (\phi, \phi)  ,\nonumber \\
&  (C_{hh\phi}, C_{h\phi\phi} )   \quad \mathrm{for}   \quad (h, \phi)  ,\nonumber \\
&  (C_{hhh}, C_{hh\phi} )   \quad \mathrm{for}   \quad (h, h)  ,\nonumber 
\end{align}
\begin{align}
4E_{i}E_j \sigma v (i j \rightarrow ) 
 = \frac{1}{16 \pi}\sqrt{\frac{s-4 m_N^2}{s}} \int \overline{|\mathcal{M}|^2} d\cos\theta  ,\\
p_{ij} = \frac{1}{2}\sqrt{\frac{\left(s-(m_{\phi_i}-m_{\phi_j})^2\right) \left(s-(m_{\phi_i}+m_{\phi_j})^2 \right)}{s}}.
\end{align}
%

\subsection{$Z'(q_1) Z'(q_2) \rightarrow N(p_1) N(p_2) $}

%
\begin{align}
&\int \sum |\mathcal{M}|^2 d\cos\theta \nonumber\\
 = & 
\frac{32 g_{B-L}^4}{m_{Z'}^4} \left(\frac{4 m_N^2 \left(m_{\phi}^2-s\right) \left(-2 m_{Z'}^2 s+4 m_{Z'}^4+s^2\right)}{\Gamma_{\phi}^2 m_{\phi}^2+\left(m_{\phi}^2-s\right)^2}-\frac{2 m_N^2 \left(4 m_N^2-s\right) \left(-4 m_{Z'}^2 s+12 m_{Z'}^4+s^2\right)}{\Gamma_{\phi}^2 m_{\phi}^2+\left(m_{\phi}^2-s\right)^2}
\right. \nonumber \\
& \left.
+\frac{2 m_N^4 \left(-8 m_{Z'}^2 s+8 m_{Z'}^4+s^2\right)+m_N^2 m_{Z'}^4 \left(4 m_{Z'}^2+s\right)-2 m_{Z'}^8}{m_N^2 \left(s-4 m_{Z'}^2\right)+m_{Z'}^4}\right) 
  \nonumber \\
& 
+\frac{32 g_{B-L}^4}{m_{Z'}^4 \sqrt{s-4 m_N^2} \sqrt{s-4 m_{Z'}^2}} \left(\frac{8 m_N^2 \left(s-m_{\phi}^2\right) \left(m_N^2 \left(-4 m_{Z'}^2 s+8 m_{Z'}^4+s^2\right)-2 m_{Z'}^6\right)}{\Gamma_{\phi}^2 m_{\phi}^2+\left(m_{\phi}^2-s\right)^2}
\right. \nonumber \\
& \left.
-\frac{4 m_N^4 s \left(s-4 m_{Z'}^2\right)+4 m_N^2 m_{Z'}^2 \left(4 m_{Z'}^2-s\right) \left(m_{Z'}^2+s\right)-m_{Z'}^4 \left(4 m_{Z'}^4+s^2\right)}{s-2 m_{Z'}^2}\right) 
\nonumber \\
&  \times \log\left( \frac{s-2 m_{Z'}^2+\sqrt{s-4 m_N^2} \sqrt{s-4 m_{Z'}^2}}{s-2 m_{Z'}^2-\sqrt{s-4 m_N^2} \sqrt{s-4 m_{Z'}^2}} \right) ,
\end{align}
\begin{align}
4E_{i}E_j \sigma v (i j \rightarrow ) 
 = \frac{1}{16 \pi}\sqrt{\frac{s-4 m_N^2}{s}} \int \overline{|\mathcal{M}|^2} d\cos\theta , \\
p_{ij} = \frac{\sqrt{s-4m_{Z'}^2}}{2}.
\end{align}
%

\section{Amplitude of the $\gamma-Z'$ scattering processes}

%
We give explicit formulas of the invariant amplitude squared.

\subsection{$ Z'(q_1) \gamma(q_2) \rightarrow  f(p_1) \bar{f}(p_2) $}

%
\begin{align}
\int \sum |\mathcal{M}|^2 d\cos\theta =& (g_{B-L} q_{Xf} e q_f)^2\frac{32}{( s-m_{Z'}^2 )^2} \nonumber \\
 & \times \left(\frac{ \sqrt{s} \left(4 m_f^2 \left(s-m_{Z'}^2\right)-8 m_f^4+m_{Z'}^4+s^2\right)}{\sqrt{s-4 m_f^2}} \log \left(\frac{\sqrt{s}+\sqrt{s-4 m_f^2}}{\sqrt{s}-\sqrt{s-4 m_f^2}}\right)  \right. \nonumber \\ 
 & \left. -s \left(4 m_f^2+s\right)-m_{Z'}^4\right) ,
\end{align}
where the IR divergence at $s=m_{Z'}^2$ is due to the on-shell $t(u)$-channel mediator and
 can be regulated by introducing thermal photon mass~\cite{Redondo:2008ec}.
\begin{align}
4E_{i}E_j \sigma v (i j \rightarrow ) 
 = \frac{1}{16 \pi}\sqrt{\frac{s-4m_f^2}{s}}
  \int \overline{|\mathcal{M}|^2} d\cos\theta , \\
p_{ij} = \frac{1}{2}\frac{s-m_{Z'}^2}{\sqrt{s}}.
\end{align}
%

\subsection{$Z'(q_1) f(q_2)\rightarrow \gamma(p_1) f(p_2)   $}

%
\begin{align}
\int \sum |\mathcal{M}|^2 d\cos\theta = & \frac{8(g_{B-L} q_{Xf} e q_f)^2}{s \left(m_f^2-s\right)^2} \left(-m_f^4 \left(m_{Z'}^2+s\right)+m_f^2 s \left(2 m_{Z'}^2+15 s\right)+m_f^6+s^2 \left(7 m_{Z'}^2+s\right) \right. \nonumber \\ 
 & \left. +\frac{2 s^2 \left(2 m_f^2 \left(m_{Z'}^2-3 s\right)-3 m_f^4-2 m_{Z'}^2 s+2 m_{Z'}^4+s^2\right) }{\sqrt{\left(s-(m_f-m_{Z'})^2\right) \left(s-(m_f+m_{Z'})^2 \right)}}\right. \nonumber \\ 
 & \left. \times \log \left(\frac{m_f^2-m_{Z'}^2+s+\sqrt{\left(s-(m_f-m_{Z'})^2\right) \left(s-(m_f+m_{Z'})^2 \right)}}{m_f^2-m_{Z'}^2+s-\sqrt{\left(s-(m_f-m_{Z'})^2\right) \left(s-(m_f+m_{Z'})^2 \right)}}\right)\right) ,
\end{align}
\begin{align}
4E_{i}E_j \sigma v (i j \rightarrow ) 
 = \frac{1}{16 \pi}  \sqrt{ \frac{s-m_f^2}{s}} \int \overline{|\mathcal{M}|^2} d\cos\theta , \\
p_{ij} = \frac{1}{2}\sqrt{\frac{\left(s-(m_f-m_{Z'})^2\right) \left(s-(m_f+m_{Z'})^2 \right)}{s}}.
\end{align}
%

\subsection{$ Z'(q_1) \bar{f}(q_2) \rightarrow \gamma(p_1) \bar{f}(p_2) $}

It is same as for $Z'(q_1) f(q_2)\rightarrow \gamma(p_1) f(p_2)$.

\bibliography{biblio}

\end{document}